\documentclass{article}
\usepackage{amssymb}
\usepackage{amsmath}
\usepackage{harvard}

\input{tcilatex}

\begin{document}

\title{\textbf{Elementary preamble to a theory of granular gases\thanks{%
The paper completes, extends, hopefully makes more perspicuous topics
covered by an earlier report of limited circulation [1]; the report itself
was an excerpt of a talk given in Bologna in 1999.}}}
\author{Gianfranco Capriz \\
Dipartimento di Matematica, Universit\`{a} di Pisa,\\
via F. Buonarroti 2, 56127 Pisa (Italy),\\
e-mail: capriz@dm.unipi.it}
\maketitle

\section{Introduction}

Granular materials partake almost dramatically at times of the properties of
solids and, under different circumstances, of some properties of gases. Some
scientists have suggested that they represent a new state of matter, to be
treated \emph{per se}; because of their ambiguity the search of an overall
continuum model for these media is quite intriguing. Even if one confines
inquiry to fast flows of sparse granules (as in granular gases, see e.g.
[2]), the range of phenomena that may occur and that should be described by
an adequate theory is vast. Within this area one is lured into adopting such
appealing terms as gross \emph{granular heat} and gross \emph{granular
temperature}, though the appropriateness of the borrowing can be challenged
as the definitions of those gross quantities involve kinetic entities, whose
detailed evolution can be in part ascertained rather than being totally
unpredictable on principle and, at most, appraised through statistics. In
any case, when exploiting vague analogies with the thermodynamics of perfect
gases, one must make allowance for the fact that, in generic granular flows,
speeds of agitation need not be, as they are in gases, many orders of
magnitude greater than the average velocity or the velocity of any
reasonably behaved observer. Thus one must be wary of a naive transfer to
each moving element in a continuum of properties inferred from experiments
within containers fixed to the walls of a laboratory; questions of
objectivity become important if not paramount. One way to avoid difficulties
would be to evaluate the relevant variables from some intrinsic local frame.
The traditional way in continuum mechanics is to appeal to gross position
and velocity gradients, a way that cannot be satisfactorily pursued in our
context. Notice that akin problems arise in the analysis of vibrations of
large and very flexible space structures (or, more down to earth, of the
motion of a misshaped soap bubble in wind): the centre of gravity need not
be superposed to a place occupied by some element of the structure; even if
perchance such superposition occurs, the movement in the vicinity of the
centre might have little to do with the gross motion of the structure; hence
the need of an alternative notion of gross rotation. A solution is easy to
find: the set of balance equations of momentum, moment of inertia, moment of
momentum can be interpreted as the tools that allow us to figure out the
background against which an objective measure of residual energy of
agitation can be achieved, with consequent satisfactory portrayal of
`thermal' concepts.

Again, the many different scales of events possible in a generic granular
flow entail a reconsideration of the ties of the concept of temperature with
the kinetic circumstances of ensembles of granules having diverse (and not
necessarily canonical) distributions of energy, thus including easily, for
instance, negative and tensorial absolute temperatures.

Finally, when one pursues a continuum model of granular flows, one must
abandon the fundamental tenet that it be possible to define, that is to
identify for ever, each material element. Then test domains of minute
diameter and the corresponding local averages take a central r\^{o}le; the
question how to average near the frontier of the body and thus how to model
boundary effects must be addressed. Moreover, in the formulation of
constitutive relations, valid either in the interior or at the boundary of
the body, local space and time-correlations might take precedence over
material space-gradients and material time-derivatives.

However, whichever changes be necessary, ultimately some alternative
paradigm should emerge and it is attractive to note, from a cursory review
of some proposals for extended versions of continuum mechanics (e.g., the
theory of hypoelasticity) or continuum thermodynamics, some stunning
similarities in the evolution equations which are arrived at. True, those
theories all presume the absence of local mesospin (which may be instead
relevant in granular flows). However, next to the equation of continuity and
Cauchy's equation an additional equation concerns the shuffling motions; it
involves a stirring tensor and rules the evolution of a Reynolds' tensor, a
symmetric tensor which both in hypoelasticity and in extended thermodynamics
coincides with momentum flux.

Precisely, the theory of hypoelasticity [3], Grad's theory of 13 moments
[4], extended thermodynamics [5], Jenkin's theory of fast flow of granular
materials [6], all suggest the addition to the classical equation of balance
of another equation which rules the evolution of the symmetric Reynolds'
tensor $H$:%
\begin{equation*}
\rho \overset{\Delta }{H}=S_{I}-div\text{ }\mathbf{s}+S_{E};
\end{equation*}%
at the same time, for Cauchy's stress tensor $T$, a very simple constitutive
law is suggested: $T=\rho H$. $\rho $ is density, $\overset{\Delta }{H}$\
the convected time derivative of $H$, $\mathbf{s}$ is the third order
stirring tensor; the tensor $S_{I}$ represents internal equilibrated
actions, and $S_{E}$ the external stir.

Here, within the mechanics of mass points, an elementary analysis which
involves predominantly velocities rather than places, is shown to lead to a
global equation of motion which suggests, by analogy, within continuum
mechanics, just the last balance equation mentioned above. Precisely that
equation may be relevant, in general, when modelling kinetic bodies (i.e.,
bodies which are in a permanent state of flux and possess no natural paragon
placement) and, in particular, granular gases.

\section{Adscititious topics in elementary mechanics}

Consider a system $\mathcal{S}$ of $N$ mass points; let $x\left( \tau
\right) $ the current centre of gravity of $\mathcal{S}$ and $\mu $\ the
total mass; $\mu ^{\left( i\right) }$, the mass of the $i-$th point; $%
x^{\left( i\right) }\left( \tau \right) $, its current place; $\tilde{y}%
^{\left( i\right) }$, its position vector with respect to the centre of
gravity in a paragon setting (e.g., the initial placement). Take for $%
R\left( \tau \right) $ a proper orthogonal tensor; split $x^{\left( i\right)
}$\ into the following sum:%
\begin{equation}
x^{\left( i\right) }\left( \tau \right) =x\left( \tau \right) +R\left( \tau
\right) s^{\left( i\right) }\text{ \ \ ; \ \ }s^{\left( i\right) }=\tilde{y}%
^{\left( i\right) }+\tilde{s}^{\left( i\right) }\left( \tau \right) 
\label{1}
\end{equation}%
involving a global rigid displacement $\left( x\left( \tau \right) ,R\left(
\tau \right) \right) $\ and an individual shuffle vector $\tilde{s}^{\left(
i\right) }$.

In a motion the velocity of each point is given by the sum%
\begin{equation}
\dot{x}^{\left( i\right) }\left( \tau \right) =\dot{x}\left( \tau \right) +%
\mathbf{e}\left( q\left( \tau \right) \otimes y^{\left( i\right) }\left(
\tau \right) \right) +R\left( \tau \right) \dot{s}^{\left( i\right) }\left(
\tau \right)   \label{2}
\end{equation}%
of speeds of entrainment (the first two addenda) and of agitation (the last
addendum): here $y^{\left( i\right) }=$\ $x^{\left( i\right) }-x$, $\mathbf{e%
}$ is Ricci's permutation tensor, $q$ is the speed of rotation so that $\dot{%
R}R^{-1}=\mathbf{e}q$.

Differentiating again (\ref{2}) with respect to time one can exhibit
explicitly entrainment, relative, Coriolis' components of acceleration:%
\begin{equation}
\ddot{x}^{\left( i\right) }=\ddot{x}+\mathbf{e}\left( \dot{q}\otimes
y^{\left( i\right) }\right) +q^{2}\left( I-c\otimes c\right) y^{\left(
i\right) }+R\ddot{s}^{\left( i\right) }+2\mathbf{e}\left( q\otimes R\dot{s}%
^{\left( i\right) }\right) ;  \label{3}
\end{equation}%
here $c$ is the unit vector associated with $q$, i.e., $q=\left\vert
q\right\vert c$, and $I$ is the identity tensor.

The total kinetic energy $\mu \kappa $ is given by%
\begin{eqnarray}
\mu \kappa &=&\frac{1}{2}\mu \dot{x}^{2}+\frac{1}{2}\mu q^{2}\left(
I-c\otimes c\right) \cdot Y+  \notag \\
&&\frac{1}{2}\dsum\limits_{i}\mu ^{\left( i\right) }\left( \dot{s}^{\left(
i\right) }\right) ^{2}+\dsum\limits_{i}\mu ^{\left( i\right) }\left( R\dot{s}%
^{\left( i\right) }\right) \cdot \mathbf{e}\left( q\otimes y^{\left(
i\right) }\right) ,  \label{4}
\end{eqnarray}%
where $\mu Y$ is Euler's inertia tensor%
\begin{equation}
\mu Y=\dsum\limits_{i}\mu ^{\left( i\right) }y^{\left( i\right) }\otimes
y^{\left( i\right) }.  \label{5}
\end{equation}

Notice that \ $\mu J$, \ $J=\left( trY\right) I-Y$ , is the usual tensor of
inertia, and that $\mu \left( I-c\otimes c\right) \cdot Y=c\cdot Jc$
coincides with the moment of inertia around a baricentric axis parallel to $c
$.

Actually, one could have started this elementary investigation taking for $x$
any moving point and enquiring subsequently how $\dot{x}\left( \tau \right) $
should be chosen so that $\frac{\mu }{2}\dot{x}^{2}$ differ the least from
the total kinetic energy; the answer would have been to choose $\dot{x}$ so
that it coincide always with the speed of the centre of gravity. Similarly,
one can choose $q$ in such a way that the speeds $\mathbf{e}\left( q\otimes
y^{\left( i\right) }\right) $ best fit the motion relative to $x\left( \tau
\right) $, in the sense that the discrepancy in kinetic energy is minimal%
\begin{equation}
\mu \kappa -\frac{1}{2}\dsum\limits_{i}\mu ^{\left( i\right) }\mathbf{e}%
\left( q\otimes y^{\left( i\right) }\right) ^{2}=\min ;  \label{6}
\end{equation}%
one needs only take for $q$ the solution of the equation%
\begin{equation}
Jq=k,  \label{7}
\end{equation}%
where $\mu k$ is the moment of momentum with respect to the centre of gravity%
\begin{equation}
\mu k=\dsum\limits_{i}\mu ^{\left( i\right) }y^{\left( i\right) }\times \dot{%
y}^{\left( i\right) }.  \label{8}
\end{equation}%
With that choice of $q$ the last term in (\ref{4}) cancels out. The kinetic
energy is split into the sum of the two, observer-dependent, usual terms as
in a rigid motion and an observer-independent term due to agitation.

To prove the statements above notice that the last term in (\ref{4}) could
be written successively as follows:%
\begin{eqnarray*}
q\cdot \dsum\limits_{i}\mu ^{\left( i\right) }y^{\left( i\right) }\otimes
\left( R\dot{s}^{\left( i\right) }\right) &=&q\cdot \left[
\dsum\limits_{i}\mu ^{\left( i\right) }\left( y^{\left( i\right) }\times 
\dot{y}^{\left( i\right) }-y^{\left( i\right) }\times \left( q\times
y^{\left( i\right) }\right) \right) \right] = \\
&=&q\cdot \left[ \dsum\limits_{i}\mu ^{\left( i\right) }y^{\left( i\right)
}\times \dot{y}^{\left( i\right) }-\mu Jq\right] ,
\end{eqnarray*}%
whereas the second addendum in the right-hand side of (\ref{4}) is exactly
the contribution to $\mu \kappa $ of a rigid rotation around $x$.

Knowledge of $\dot{x}$ and $q$ (and hence of $x$ and $R$) offers the chance
to fix a background against which the motion of agitation can be measured
objectively. To determine $\dot{x}$ and $q$, appeal can be made to the
global equations%
\begin{equation}
\mu \ddot{x}=f\text{ \ \ , \ \ }\mu \dot{k}=m,  \label{9}
\end{equation}%
where $f$ and $m$ are resultant and resultant moment of external forces on $%
\mathcal{S}$. Success is obvious if there is no agitation and the motion is
trivially rigid. However, in general one is not so fortunate because there
may be an influence of shuffle and agitation certainly on $J$ and,
perchance, on $f$ and $m$; notice, in particular, that $J$ depends
critically on expansion/contraction of $\mathcal{S}$ because%
\begin{equation}
\dot{J}=\left( tr\dot{Y}\right) I-\dot{Y}  \label{10}
\end{equation}%
and%
\begin{equation}
\dot{Y}=2symK,  \label{11}
\end{equation}%
where%
\begin{equation}
\mu K=\dsum\limits_{i}\mu ^{\left( i\right) }y^{\left( i\right) }\otimes 
\dot{y}^{\left( i\right) }  \label{12}
\end{equation}%
is the tensor moment of momentum, the evolution of which must be known to
evaluate the evolution of $Y$. In a rigid motion $K$ depends on $Y$ and $q$
only, $K=-\left( \mathbf{e}q\right) Y$, and no further developments are
required. Instead, in general, as remarked above, a deeper analysis of the
motions of $\mathcal{S}$ needs be effected in advance. We need to modify the
developments at the beginning of this section as a premise; rather than (\ref%
{1}) take now%
\begin{equation}
x^{\left( i\right) }\left( \tau \right) =x\left( \tau \right) +G\left( \tau
\right) s^{\left( i\right) }\left( \tau \right) ,  \label{13}
\end{equation}%
where $G$ is a tensor with positive determinant; correspondingly,%
\begin{equation}
\dot{x}^{\left( i\right) }=\dot{x}+By^{\left( i\right) }+G\dot{s}^{\left(
i\right) }\text{ \ \ , \ \ }B=\overset{\cdot }{G}G^{-1},  \label{14}
\end{equation}%
and%
\begin{equation}
\ddot{x}^{\left( i\right) }=\ddot{x}+\left( \dot{B}+B^{2}\right) y^{\left(
i\right) }+2BG\dot{s}^{\left( i\right) }+G\ddot{s}^{\left( i\right) },
\label{15}
\end{equation}%
if $y^{\left( i\right) }$\ is equal to $Gs^{\left( i\right) }$.

Thus, shuffle and agitation are now supposed to be remainders beyond an
affine, rather than rigid, motion. The kinetic energy is given by the sum%
\begin{equation}
\frac{1}{2}\mu \dot{x}^{2}+\frac{1}{2}\mu tr\left( BYB^{T}\right) +\frac{1}{2%
}\dsum\limits_{i}\mu ^{\left( i\right) }\left( G\dot{s}^{\left( i\right)
}\right) ^{2}+\dsum\limits_{i}\mu ^{\left( i\right) }\left( G\dot{s}^{\left(
i\right) }\right) \cdot \left( By^{\left( i\right) }\right) .  \label{16}
\end{equation}%
However, if $B$ is chosen so as to make $\mu \kappa -\dsum\limits_{i}\mu
^{\left( i\right) }\left( By^{\left( i\right) }\right) ^{2}$ a minimum,
i.e., as to satisfy the condition%
\begin{equation}
BY=K^{T},  \label{17}
\end{equation}%
the last term in (\ref{16}) vanishes; in fact, even the mixed kinetic tensor%
\begin{equation}
\dsum\limits_{i}\mu ^{\left( i\right) }\dot{s}^{\left( i\right) }\otimes
By^{\left( i\right) }  \label{18}
\end{equation}%
vanishes, not only its trace, so that a relatively compact expression is
available also for the kinetic energy tensor $\mu W=\frac{1}{2}%
\dsum\limits_{i}\mu ^{\left( i\right) }\dot{x}^{\left( i\right) }\otimes 
\dot{x}^{\left( i\right) }$, as follows:%
\begin{equation}
\mu W=\frac{1}{2}\mu \dot{x}\otimes \dot{x}+\frac{1}{2}\mu BYB^{T}+\frac{1}{2%
}\mu H.  \label{19}
\end{equation}%
where, if $H_{\ast }$ is Reynolds's kinetic tensor of agitation, $H$\ is its
transformed expression by $G$:%
\begin{equation}
\mu H_{\ast }=\dsum\limits_{i}\mu ^{\left( i\right) }\dot{s}^{\left(
i\right) }\otimes \dot{s}^{\left( i\right) }\text{ \ \ , \ \ }H=GH_{\ast
}G^{T}.  \label{20}
\end{equation}

It is important to notice, in particular for later developments, that all
vectors $\dot{s}^{\left( i\right) }$ and the tensor $H_{\ast }$ are
absolute, i.e., not affected by changes or movements of the observer.

To determine $\dot{x}$\ and $B$, appeal is now made to the first one of (\ref%
{9}) and to the equation of balance of the tensor moment of momentum%
\begin{equation}
\mu \left( \dot{K}-BK-H\right) =M-A,  \label{21}
\end{equation}%
where $M$ and $A$ are respectively the tensor moment of external and
internal forces acting on $\mathcal{S}$%
\begin{equation}
M=\dsum\limits_{i}y^{\left( i\right) }\otimes f^{ext\left( i\right) }\text{
\ \ , \ \ }A=-\dsum\limits_{i}y^{\left( i\right) }\otimes f^{int\left(
i\right) }.  \label{22}
\end{equation}%
Because the vector moment of internal forces vanishes, $A$ is a symmetric
tensor. Remark that (\ref{11}) can be read as the assertion that the
`reference' tensor of inertia $Y_{\ast }=G^{-1}YG^{-T}$ is constant
irrespective of shuffle.

We still need the evolution equation for $H$; it is easily obtained from
Newton's law, multiplying both members tensorially by $\dot{s}^{\left(
i\right) }$, summing and transforming through $G$:%
\begin{equation}
\mu \left( \dot{H}+BH+HB^{T}\right) =S-Z;  \label{23}
\end{equation}%
here $S$ is the stirring tensor of external forces%
\begin{equation}
S=2sym\dsum\limits_{i}\left( G\dot{s}^{\left( i\right) }\right) \otimes
f^{ext\left( i\right) },  \label{24}
\end{equation}%
and a similar definition but with opposite sign applies to $Z$, involving
internal forces. The queer choice of sign for $A$ and $Z$ has to do with a
convention appropriate in a distinct, later context. Finally, if one remarks
that%
\begin{equation}
K=symK+\frac{1}{2}\mathbf{e}k\text{ \ \ , \ \ }M=symM+\frac{1}{2}\mathbf{e}m,
\label{25}
\end{equation}%
the system of equations to explore becomes%
\begin{equation}
\left\{ 
\begin{array}{c}
\mu \ddot{x}=f, \\ 
\dot{k}=m, \\ 
\mu sym\left( \dot{K}-BK\right) =symM+\mu H-A, \\ 
\mu \left( \dot{H}+BH+HB^{T}\right) =S-Z.%
\end{array}%
\right.   \label{26}
\end{equation}

Strictly, the first two equations bear upon the preferred reference, if only
its rotational speed $q$ is defined through the relation%
\begin{equation*}
q=J^{-1}k,
\end{equation*}%
whereas the last two lead to global hints on agitation; but, generally, the
two sets are strongly linked. Actually, for some later purposes, it is more
convenient to keep together the two equations for the symmetric and skew
components of $K$ and to add to the list the evolution equation for $Y$:%
\begin{equation}
\left\{ 
\begin{array}{c}
\mu \ddot{x}=f, \\ 
\mu \left( \dot{K}-BK\right) =M-A+\mu H, \\ 
\dot{Y}=YB^{T}+BY, \\ 
\mu \left( \dot{H}+BH+HB^{T}\right) =S-Z.%
\end{array}%
\right.  \label{27}
\end{equation}

Notice, in the second and fourth equation, that the quantities between
brackets express the convected time derivative $\overset{\Delta }{K}$\ of $K$%
\ and $\overset{\Delta }{H}$\ of $H$, based on the spin tensor $B$.

If all $\dot{s}^{\left( i\right) }$\ vanish, then all $\dot{s}^{\left(
i\right) }$\ maintain their initial values, $H$ vanishes and the local
motion is affine, or pseudo-rigid in the therminology of Cohen and Muncaster
[7]. Their results could be borrowed here; one needs to study only the
reduced system of the first three equations of (\ref{27}) in $x$ and $B$.

More generally, if $f$, $M$, $A$, $Z$ and $S$ depend at most on $x$, $\dot{x}
$, $G$, $B$ and $H$, then (\ref{26}) can be interpreted as a differential
system in $x$, $G$, and $H$.

Circumstances could be called \emph{kinetic} when the system has no
physically relevant paragon setting, its behaviour is ruled by abrupt
responses to current circumstances and, as a consequence, in our model, $f$, 
$M$, and $S$ need depend at most on $v=\dot{x}$, $B$ and $H$; then (\ref{27}%
) becomes a first order system in $v$, $B$, $Y$ and $H$. This system would
merit scrutiny on its own, perhaps under appropriate, special choices of $f$%
, $M$, $A$, $Z$ and $S$.

A case of special interest is met when $A=\mu H$ or, at least the quantity $%
\left( A-\mu H\right) $ together with $f$ and $M$ depend at most on $v$, $B$
and $Y$; then the first three equations can be dealt separately from the
last.

More subtle is the case in when $f$, $M$ and $S$ depend also on $\kappa
+\varphi _{I}$ where $\varphi _{I}$ is the potential energy of internal
forces, supposing that they all be conservative; or, rather, when they
depend also on the manner the mass points of $\mathcal{S}$ can be classed in
families of increasing total energy, or, more deeply, in families of
approximately equal Reynolds tensor.

\section{The kinetic energy theorem and some corollaries}

A tensor kinetic energy theorem can be derived easily from (\ref{27}) by
adding term by term first, second and fourth equation after tensorial
multiplication of the first by $\dot{x}$, of the second by $B$, and after
multiplication by $\frac{1}{2}$ of the fourth; finally, by taking the
symmetric parts of all terms:%
\begin{eqnarray}
\mu \dot{W} &=&sym\left[ \mu \ddot{x}\otimes \dot{x}+\mu \left( \dot{B}%
YB^{T}+B^{2}YB^{T}\right) \right] +\frac{1}{2}\mu \dot{H}=  \notag \\
&=&sym\left[ \mu \ddot{x}\otimes f+B\left( M-A\right) \right] +\frac{1}{2}%
\left( S-Z\right) .  \label{28}
\end{eqnarray}%
More particularly, but also with deeper meaning, if one takes the trace, one
arrives at the more usual kinetic energy theorem%
\begin{equation}
\dot{\kappa}=\dot{x}\cdot f+\left( M-A\right) \cdot B^{T}+\frac{1}{2}%
tr\left( S-Z\right) .  \label{29}
\end{equation}

Remark that%
\begin{equation}
\begin{array}{c}
M\cdot B^{T}+\frac{1}{2}trS=\dsum\limits_{i}\dot{y}^{i}\cdot f_{E}^{i}, \\ 
-\left( A\cdot B^{T}+\frac{1}{2}trZ\right) =\dsum\limits_{i}\dot{y}^{i}\cdot
f_{I}^{i}.%
\end{array}
\label{30}
\end{equation}%
The standard requirement that the power of internal actions be invariant for
any rigid change of speed, when $B$ is an arbitrary \ skew tensor, also
leads to the condition%
\begin{equation*}
A\in Sym,
\end{equation*}%
which was already noticed on equivalent grounds.

If potentials $\varphi _{E}$, $\varphi _{I}$\ exist for external and
internal forces, respectively, then%
\begin{equation}
f_{E}^{i}=\frac{\partial \varphi _{E}}{\partial x^{i}}\text{ \ \ , \ \ }%
f_{I}^{i}=\frac{\partial \varphi _{I}}{\partial x^{i}},  \label{31}
\end{equation}%
a theorem of energy conservation follows:%
\begin{equation}
\kappa +\varphi =const\text{ \ \ \ , \ \ \ }\varphi =\varphi _{E}+\varphi
_{I}.  \label{32}
\end{equation}%
Notice that, on the one hand, for (\ref{14}),%
\begin{equation*}
\dot{\varphi}=\dsum\limits_{i}\frac{\partial \varphi }{\partial x^{\left(
i\right) }}\cdot \left( \dot{x}+By^{\left( i\right) }+G\dot{s}^{\left(
i\right) }\right) ,
\end{equation*}%
and, on the other hand,%
\begin{equation*}
\frac{\partial \varphi }{\partial x}=\dsum\limits_{i}\frac{\partial \varphi 
}{\partial x^{\left( i\right) }}\text{ \ \ \ , \ \ \ }\frac{\partial \varphi 
}{\partial G}=\dsum\limits_{i}\frac{\partial \varphi }{\partial x^{\left(
i\right) }}\otimes s^{\left( i\right) },
\end{equation*}%
so that%
\begin{equation*}
\dot{\varphi}=\dot{x}\cdot \left( \dsum\limits_{i}\frac{\partial \varphi }{%
\partial x^{\left( i\right) }}\right) +\overset{\cdot }{G}\cdot \left(
\dsum\limits_{i}\frac{\partial \varphi }{\partial x^{\left( i\right) }}%
\otimes s^{\left( i\right) }\right) +G\cdot \left( \dsum\limits_{i}\frac{%
\partial \varphi }{\partial x^{\left( i\right) }}\otimes \dot{s}^{\left(
i\right) }\right) .
\end{equation*}%
Juxstaposition with the right-hand side of (\ref{29}) suggests the
`constitutive laws'%
\begin{equation*}
f=\frac{\partial \varphi }{\partial x}\text{ \ \ \ , \ \ \ \ }M-A=\frac{%
\partial \varphi }{\partial G}G^{T}\text{ \ \ \ , \ \ \ }S-Z=2sym\left(
\dsum\limits_{i}\dot{s}^{\left( i\right) }\otimes \frac{\partial \varphi }{%
\partial s^{\left( i\right) }}\right) ,
\end{equation*}%
or, more precisely,%
\begin{equation*}
f=\frac{\partial \varphi _{E}}{\partial x}\text{ \ \ \ , \ \ \ \ }M=\frac{%
\partial \varphi _{E}}{\partial G}G^{T}\text{ \ \ \ , \ \ \ \ }A=-\frac{%
\partial \varphi _{I}}{\partial G}G^{T},
\end{equation*}%
\begin{equation}
S=2sym\left( \dsum\limits_{i}\dot{s}^{\left( i\right) }\otimes \frac{%
\partial \varphi _{E}}{\partial s^{\left( i\right) }}\right) \text{ \ \ \ ,
\ \ \ \ }Z=-2sym\left( \dsum\limits_{i}\dot{s}^{\left( i\right) }\otimes 
\frac{\partial \varphi _{I}}{\partial s^{\left( i\right) }}\right) .
\label{33}
\end{equation}%
$\varphi _{E}$ may be influenced also by $x$ and $skwG$, besides $symG$ and $%
\dot{s}^{\left( i\right) }$; on the contrary, $\varphi _{I}$ must be
observer independent, hence it must not involve $x$ and may depend on $G$
only through the product $C=G^{T}G$, so that%
\begin{equation}
A=-2G\frac{\partial \varphi _{I}}{\partial C}G^{T}.  \label{34}
\end{equation}%
Of course, in general, $A$ and $Z$ have also dissipative components beside
the conservative components expressed, as above, through $\varphi _{I}$.

Sometimes there is an interest for a `reduced' theorem involving the `gross'
kinetic energy tensor $\tilde{W}$ per unit mass%
\begin{equation*}
\tilde{W}=\frac{1}{2}\dot{x}\otimes \dot{x}+\frac{1}{2}BYB^{T}.
\end{equation*}%
The theorem is signified by the reduced equation%
\begin{equation*}
\mu \overset{\cdot }{\tilde{W}}=sym\left[ \mu \dot{x}\otimes f+B\left(
M-A\right) \right] ,
\end{equation*}%
and leads, by difference from (\ref{28}) (and when the trace is taken), to a
`principle' of energy balance. The latter is, in the present context,
nothing else but a corollary of the last equation (\ref{27}) but with a
different promotion of terms, the leading r\^{o}le being played by the
`internal energy'%
\begin{equation}
\varepsilon =\frac{1}{2}trH-\varphi _{I}.  \label{35}
\end{equation}

As we can avail ourselves of the more powerful relation (\ref{28}), we do
not pursue here the consequences of that corollary. Actually, for later
purposes, one can write the last equation (\ref{27}) in an equivalent form
which approaches the principle of conservation more closely, using a tensor
of energy%
\begin{equation*}
E=\frac{1}{2}H+\frac{1}{3}\varphi _{I}I;
\end{equation*}%
precisely%
\begin{equation*}
\mu \left( \dot{E}+BE+EB^{T}\right) =\frac{1}{2}\left( S-\hat{Z}\right) ,
\end{equation*}%
where%
\begin{eqnarray*}
\hat{Z} &=&Z-\frac{1}{3}\dot{\varphi}_{I}I-\frac{2}{3}\varphi _{I}symB= \\
&=&-2sym\left[ \dsum\limits_{i}\dot{s}^{\left( i\right) }\otimes \frac{%
\partial \varphi _{I}}{\partial s^{\left( i\right) }}-\frac{1}{6}\dot{\varphi%
}_{I}I-\frac{1}{3}\varphi _{I}symB\right] .
\end{eqnarray*}

\section{Energy distributions}

The topics of this section are textbook affairs; because of the interest
here in non-canonical instances, they are recalled nonetheless in essence
and with the appropriate slant.

\subsection{The scalar case}

When the mass points are very numerous, though with bounded total mass $\mu $%
, the `averages' $x$, $k$, $Y$, etc. acquire prominent import. At the same
time, as mentioned at the end of the previous section, the resultant actions
on $\mathcal{S}$, expressed by $f$, $M$, $S$, may come to depend (not only
on kinematic variables such as $v$, $B$, $Y$, $H$, but, as hinted, also) on
the way mass-points can be parcelled out in families, each family comprising
points with an energy of agitation falling within a limited range, say $%
\left[ \left( j-1\right) \mu \varepsilon \delta ,j\mu \varepsilon \delta %
\right] $, $j=1,2...$ (here $\delta $ is a positive number). The fraction $%
\frac{N_{i}}{N}$ of mass points belonging to the $j-$th family is measured
by a non-negative constant $\gamma _{j}$.

Then an histogram can be drawn as a graph of a piecewise constant function $%
\bar{\gamma}\left( \xi \right) $ having the value $\gamma _{j}$ for $\xi $
within the interval $\left[ \left( j-1\right) \delta ,j\delta \right] $.
Notice that there will always be a value, say $J$ of the index, such that $%
N_{J}>0$, whereas all $N_{j}$ with index larger than $J$ vanish; that it is
so because the total energy $m\varepsilon $ is (approximately equal to and)
not less than $\mu \varepsilon \delta \dsum\limits_{j}\left( j-1\right) N_{j}
$; that each term in the sum cannot exceed $\delta ^{-1}$; that any non-null
value of $N_{j}$ is a positive integer, hence no less than 1; and, in
conclusion, that $J-1\leq \delta ^{-1}$.

The function $\bar{\gamma}\left( \xi \right) $ satisfies a normalization
condition%
\begin{equation*}
\int_{0}^{\infty }\bar{\gamma}\left( \xi \right) d\xi =\delta
\dsum\limits_{j}\gamma _{j-1}=\frac{1}{N}\dsum\limits_{j}N_{j}=1.
\end{equation*}

A second important relation is derived easily; remark that%
\begin{equation*}
\int_{0}^{\infty }\xi \bar{\gamma}\left( \xi \right) d\xi
=\dsum\limits_{j}\gamma _{j}\int_{\left( j-1\right) \delta }^{j\delta }\xi
d\xi =\dsum\limits_{j}j\gamma _{j}\delta ^{2}-\frac{1}{2}\delta ,
\end{equation*}%
and that, on the other hand,%
\begin{equation*}
\dsum\limits_{j}\mu \varepsilon \delta ^{2}\left( j-1\right) \gamma _{j}\leq
\mu \varepsilon \leq \dsum\limits_{j}\mu \varepsilon \delta ^{2}j\gamma _{j}
\end{equation*}%
or%
\begin{equation*}
1\leq \delta ^{2}\dsum\limits_{j}j\gamma _{j}\leq 1+\delta .
\end{equation*}

In conclusion,%
\begin{equation*}
1-\frac{\delta }{2}\leq \int_{0}^{\infty }\xi \bar{\gamma}\left( \xi \right)
d\xi \leq 1+\frac{\delta }{2}.
\end{equation*}

Our analysis becomes more fluent if we proceed to smooth out the histogram
(perhaps imagining that the range of $\delta $\ are taken ever smaller) to
become the graph of a (continuous, even smooth) function $\gamma \left( \xi
\right) $ with the properties that $\gamma \left( \xi \right) d\xi $ gives
the fraction of mass points with energy of agitation within the interval $%
\left( \mu \varepsilon \xi ,\mu \varepsilon \left( \xi +d\xi \right) \right) 
$ and that the following normalization conditions apply%
\begin{equation*}
\int_{0}^{\infty }\gamma \left( \xi \right) d\xi =1\text{ \ \ , \ \ \ }%
\int_{0}^{\infty }\xi \gamma \left( \xi \right) d\xi =1.
\end{equation*}

It is easy to contrive distribution functions satisfying all conditions
noticed so far for $\gamma $. Take any function $\lambda \left( \bar{\xi}%
\right) $ defined over $\left[ 0,+\infty \right) $ with non-negative, not
everywhere null values (even a measure) and integrable, together with $\bar{%
\xi}\lambda \left( \bar{\xi}\right) $, over $\left[ 0,+\infty \right) $%
\begin{equation*}
\int_{0}^{\infty }\lambda \left( \bar{\xi}\right) d\bar{\xi}=\rho \text{ \ \
, \ \ \ }\int_{0}^{\infty }\bar{\xi}\lambda \left( \bar{\xi}\right) d\bar{\xi%
}=\sigma ;\text{ \ \ \ \ }0<\rho ,\sigma <\infty .
\end{equation*}

By choosing%
\begin{equation*}
\gamma \left( \xi \right) =\frac{\sigma }{\rho ^{2}}\lambda \left( \frac{%
\rho \xi }{\sigma }\right) ,
\end{equation*}%
one obtains just one of the desired functions.

It is an easy matter to check that the following choices for $\gamma $
satisfy all requirements mentioned above. It is appropriate to emphasize
that the abscissa $\xi $ for the histogram is chosen here as to be
non-dimensional and such that mass points for which $\xi =1$ have energy per
unit mass exactly equal to the total energy per unit mass. The use of
non-dimensional variables may give an impression of excessive
specialization; in the formulae, in fact, the choice of constants is
mandatory:

\begin{description}
\item[(i)] Canonical:%
\begin{equation}
\gamma =e^{-\xi }\text{ \ \ , \ \ \ for all }\xi .  \label{36}
\end{equation}%
For some systems, such as monoatomic gases, this distribution is requisite
under `quasi-static' conditions.

\item[(ii)] Power law:%
\begin{equation}
\gamma =24\left( 2+\xi \right) ^{-4}\text{ \ \ , \ \ \ for all }\xi .
\label{37}
\end{equation}%
Remark that other negative powers different from $-4$, strictly less than $-2
$ could be contemplated; the numerical factors would then be obviously
different.

\item[(iii)] Piece-wise constant: For any constant $\beta \in \left[
0,1\right) $, take%
\begin{equation*}
\gamma \left( \xi \right) =0,\text{ \ \ when \ \ }0\leq \xi <\beta \text{ \
\ or \ \ }\xi >2-\beta ,
\end{equation*}%
\begin{equation}
\gamma \left( \xi \right) =\frac{1}{2\left( 1-\beta \right) },\text{ \ \
when \ \ }\beta <\xi <2-\beta .  \label{38}
\end{equation}%
The limit for $\beta \rightarrow 1$ is a measure with an atom at $\xi =1$.

\item[(iv)] Piece-wise linear: For any $\beta \in \left[ \frac{3}{2},3\right]
$, take%
\begin{equation*}
\gamma \left( \xi \right) =0,\text{ \ \ when \ \ }\xi >\beta ,
\end{equation*}%
\begin{equation}
\gamma \left( \xi \right) =2\beta ^{-3}\left[ 3\left( 2-\beta \right) \xi
+\left( 2\beta -3\right) \beta \right] ,\text{ \ \ when \ \ }0\leq \xi \leq
\beta .  \label{39}
\end{equation}

\item[(v)] Piece-wise exponential: Given the constants $\beta $, positive; $%
\xi _{1}\geq 0$, $\xi _{2}>\xi _{1}$; and $\alpha $ real; take%
\begin{equation*}
\gamma \left( \xi \right) =0,\text{ \ \ when \ \ }0\leq \xi \leq \xi _{1}%
\text{ \ \ or \ \ }\xi >\xi _{2},
\end{equation*}%
\begin{equation*}
\gamma \left( \xi \right) =\beta e^{-\alpha \xi },\text{ \ \ when \ \ }\xi
\in \left[ \xi _{1},\xi _{2}\right] ,
\end{equation*}%
where%
\begin{equation*}
\beta =\frac{\alpha }{e^{\alpha \xi _{2}}-e^{-\alpha \xi _{1}}}
\end{equation*}%
and $\alpha $ is a solution of the equation%
\begin{equation}
\alpha =\frac{\left( 1+\alpha \xi _{2}\right) e^{-\alpha \xi _{2}}-\left(
1+\alpha \xi _{1}\right) e^{-\alpha \xi _{1}}}{e^{-\alpha \xi
_{2}}-e^{-\alpha \xi _{1}}}.  \label{40}
\end{equation}%
To lighten developments, we refer below only to the choice $\xi _{1}=0$, when%
\begin{equation}
\gamma \left( \xi \right) =\frac{\alpha }{e^{\alpha \xi _{2}}-1}e^{-\alpha
\xi },  \label{41}
\end{equation}%
where $\alpha $ satisfies the equation%
\begin{equation}
\alpha -1=\frac{\alpha \xi _{2}}{1-e^{-\alpha \xi _{2}}}.  \label{42}
\end{equation}%
An inspection of (\ref{42}) requiring only accurate evaluations of orders of
magnitude shows that: There is one and only one value of $\alpha $
satisfying (\ref{42}) for each choice of $\xi _{2}$ larger than 1. There are
no values of $\alpha $ satisfying (\ref{42}) for $\xi _{2}<1$. The function $%
\alpha \left( \xi _{2}\right) $, thus defined, is strictly increasing from $%
-\infty $ to 1; it vanishes for $\xi _{2}=2$; its approximate expression in
the neighborhood of 2 is%
\begin{equation*}
\alpha \simeq \frac{3}{2}\left( \xi _{2}-2\right) .
\end{equation*}%
When $\xi _{2}$ tends to 1 and $\alpha $ to $-\infty $, the distribution (%
\ref{41}) approaches a $\delta -$ function with an atom at $\xi _{2}=1$.
When $\xi _{2}$ tends to the value 2 and $\alpha $ to zero, the distribution
(\ref{41}) tends to be piece-wise constant: $\gamma \left( \xi \right) =%
\frac{1}{2}$ for $0<\xi <2$ and null otherwise. When $\xi _{2}$ tends to $%
\infty $ and $\alpha $ to 1, the distribution tends to be canonical. This
distribution is reminiscent of one which is suitable for quantum systems
though allowing in that case only a finite number of states. As is well
known, for them, with appropriate care, a sudden transit from a distribution
with $\alpha $ positive into one with $\alpha $ negative can be achieved
experimentally.

\item[(vi)] Sinusoidal: Given a constant $\alpha $ less than $\frac{\pi ^{2}%
}{4}$ take%
\begin{equation*}
\gamma \left( \xi \right) =0,\text{ \ \ when \ \ }\xi >2\left( 1-\frac{%
4\alpha }{\pi ^{2}}\right) ^{-1},
\end{equation*}%
\begin{equation}
\gamma \left( \xi \right) =\frac{1}{2}\left( 1-\frac{4\alpha }{\pi ^{2}}%
\right) \left( 1+\alpha \cos \frac{\pi }{2}\left( 1-\frac{4\alpha }{\pi ^{2}}%
\right) \xi \right) .  \label{43}
\end{equation}%
The limit when $\alpha $ goes to zero coincides with the limit case (iii)
when $\beta \rightarrow 0$. $\gamma $ either decreases or increases with
increasing $\xi $, depending on the sign of $\alpha $. This distribution may
be of interest when energy and numerosity are functions of an angle from a
given direction.

\item[(vii)] Fermi and Bose. For $\beta $ non-null%
\begin{equation}
\gamma \left( \xi \right) =\left\vert e^{\left\vert \beta \right\vert \xi
}\left( e^{\beta }-1\right) ^{-1}-1\right\vert ^{-1}  \label{43bis}
\end{equation}%
satisfies the first normalization condition; now choose $\beta $\ so as to
fulfill also the second one, which requires that%
\begin{equation}
\beta ^{2}=\text{Di}\log \left( e^{\beta }-1\right) \text{,}  \label{43ter}
\end{equation}%
where%
\begin{equation}
\text{Di}\log y=\dsum\limits_{k=1}^{\infty }\frac{y^{k}}{k^{2}}=\int_{y}^{0}%
\frac{\lg \left( 1-t\right) }{t}dt.  \label{43quater}
\end{equation}%
There are two solutions, one with $\beta $\ negative (Bose-Einstein subcase)
and the other with $\beta $ positive (Fermi-Dirac subcase):%
\begin{equation}
\beta =-0.814651...\text{ \ \ \ , \ \ \ \ \ }\beta =1.405050...
\label{43penta}
\end{equation}
\end{description}

\subsection{The tensor case}

As hinted repeatedly, a classification of mass points in families with
specific energy is inadequate at times to cast satisfactory constitutive
laws completing system (\ref{26}) or (\ref{27}). One can explore broader
alternatives (see, e.g., [8]), in particular one can attempt the
classification in tensorial terms; i.e., when $\varphi _{I}$ vanishes, in
terms of the Reynolds tensor. Precisely, to pursue here the latter (special,
but most important) case, think of the linear space of symmetric tensors
and, within it, the manifold $\mathcal{N}$ of all rank-1 tensors, i.e., of
all tensors $N$\ of the type $v\otimes v$ with $v$ any vector. With $\gamma $%
\ a function of the tensor variable $H^{-\frac{1}{2}}NH^{-\frac{1}{2}}$,\
call now%
\begin{equation*}
\gamma \left( H^{-\frac{1}{2}}NH^{-\frac{1}{2}}\right) \left( H^{-1}\cdot
dN\right) 
\end{equation*}%
the number fraction of mass points with Reynolds tensor in the immediate
neighborhood of $N$; imagine again the histogram of $\gamma $ smoothed out
to render it continuous and differentiable. It would also be integrable over 
$\mathcal{N}$ together with its product by $H^{-1}N$ and with the properties
of normality%
\begin{equation*}
\int_{\mathcal{N}}\gamma \left( H^{-\frac{1}{2}}NH^{-\frac{1}{2}}\right)
H^{-1}\cdot dN=1,
\end{equation*}%
\begin{equation*}
\int_{\mathcal{N}}\left( H^{-1}\cdot N\right) \gamma \left( H^{-\frac{1}{2}%
}NH^{-\frac{1}{2}}\right) \left( H^{-1}\cdot dN\right) =1.
\end{equation*}%
Then, formally, one can proceed in analogy with the scalar case; in
particular, one can introduce the extended canonical distribution, valid on $%
\mathcal{N},$%
\begin{equation*}
\gamma =e^{-H^{-1}\cdot N}
\end{equation*}%
(see, e.g., [9]) and explore also alternative distributions.

\section{Granular temperature}

When the intimation at the end of Section 3 is mandatory, knowledge of the
distribution $\gamma $ becomes essential; then the need would appear, from
the developments of Section 4, to seek another equation (a sort of Boltzmann
equation) to describe the evolution of $\gamma $. Actually it often occurs
that either the class of the accessible distributions is restricted on
physical grounds and each distribution is then identified by few parameters,
or, at least, only a few variables linked to the distribution are essential
and suffice. Such is the granular temperature $\vartheta $, which is defined
below:%
\begin{equation}
\vartheta =-\frac{d\xi }{d\left( \log \gamma \right) }=-\gamma \frac{d\xi }{%
d\gamma }.  \label{44}
\end{equation}%
It is a constant parameter in case (i). If the case (iii) is taken as the
limit case (v) for $\alpha \rightarrow 0$, the associated temperature is
infinite. In case (v), subcase $\xi _{1}=0$, then $\vartheta =\alpha ^{-1}$;
hence the graph of temperature against $\xi _{2}$ shows $\vartheta $
decreasing from 0 to $-\infty $ as $\xi _{2}$ ranges from $1$ to $2$, and
then decreasing from $+\infty $ to 1 as $\xi _{2}$ grows to $+\infty $. It
is noteworthy to remark that null temperature can be achieved only through
negative values; transition from positive to negative temperature can occur
only through $\infty $, as is known experimentally.

In distributions different from exponential, as in case (ii), the derivative
(\ref{44}) is not a constant; one can declare that temperature is
meaningless for such cases or be satisfied with an average value $\vartheta
_{w}$ of the derivative over its support.

More precisely, suppose that, as required for invertibility, $\gamma \left(
\xi \right) $ be strictly monotone and positive only over an interval $%
\left( \xi _{1},\xi _{2}\right) \in \left( \xi _{1},+\infty \right] $; then
for the weaker definition $\vartheta _{w}$ of the temperature, one obtains%
\begin{eqnarray}
\vartheta _{w} &=&\left( \gamma \left( \xi _{1}\right) -\gamma \left( \xi
_{2}\right) \right) ^{-1}\int_{\gamma \left( \xi _{1}\right) }^{\gamma
\left( \xi _{2}\right) }\frac{d\xi }{d\left( \log \gamma \right) }d\gamma = 
\notag \\
&=&\left( \gamma \left( \xi _{1}\right) -\gamma \left( \xi _{2}\right)
\right) ^{-1}\int_{\xi _{1}}^{\xi _{2}}\gamma d\xi =\frac{1}{\gamma \left(
\xi _{1}\right) -\gamma \left( \xi _{2}\right) }\text{.}  \label{45}
\end{eqnarray}

Summing up, for the special distributions above, the temperature turns out
to be%
\begin{equation*}
\text{(i) \ \ }\vartheta =1\text{ \ \ ; \ \ (ii) \ \ }\vartheta _{w}=\frac{3%
}{2}\text{ \ \ ; \ \ (iii) \ \ }\vartheta =\infty \text{ ;}
\end{equation*}%
\begin{equation*}
\text{(iv) \ \ }\vartheta _{w}=\frac{1}{6}\frac{\beta ^{2}}{\beta -2}\text{
\ ; \ \ \ (vi) \ }\vartheta _{w}=\frac{\pi ^{2}}{\left( \pi ^{2}-4\alpha
\right) \alpha }\text{ \ ; \ \ \ (vi)\ \ }\vartheta _{w}=\left\vert \frac{%
2-e^{\beta }}{e^{\beta }-1}\right\vert .
\end{equation*}%
In case (v), subcase $\xi _{1}=0$, $\vartheta =\alpha ^{-1}$, hence $%
\vartheta $ varies with the choice of $\xi _{2}$ as was described above.

As hinted at the end of Section 2, the temperature (or its reciprocal, the
temperance), even when allowed to roam over the entire real axis, may be yet
inadequate to allow closure of system (\ref{26}), (\ref{27}) for a
sufficiently wide class of bodies. Then one can explore broader alternatives
(see, e.g., [8]) and insinuate the idea that a tensor related to the
distribution of shufflings may be of help. Precisely, one may think to the
tensor case discussed in Section 4. Then, over the support of $\gamma $ on $%
\mathcal{N}$ also $\log \gamma $ becomes meaningful and the following \emph{%
temperature tensor} can be defined:%
\begin{equation}
-\frac{dN}{d\left( \log \gamma \right) }.  \label{46}
\end{equation}

As already mentioned, by far-fetched analogy with standard cases, one may
dream up occurrences when $H$, $S$, and $Z$ are functions of the temperature
tensor on physical grounds; then, the last equation (\ref{27}) becomes an
evolution equation for the temperature tensor. Naturally such dreams need by
substantiated by offering some concrete cases where the fantasy bears
fruits; for the moment, let rest us content to have drawn attention to a
possible avenue which is wider than that opened up by the mere scalar
temperature and hence perhaps more attuned to the study of granular
materials.

\section{Transition to a continuum model}

The system (\ref{27}) is already written in such a way as to suggest the
possible evolution equations for a continuum, where the image of the body is
a fit region rather than a discrete set in Euclidean space.

Now mass density $\rho $ takes the defining r\^{o}le for mass; $\rho $ and
the current place $x$, the current gross local shape $G$, the current tensor
of moment of inertia $Y$, and the kinetic tensor $H$, are all fields.

The conservation equation for mass is the classical one%
\begin{equation}
\dot{\rho}+\rho div\dot{x}=0,  \label{47}
\end{equation}%
and its twin for tensor moment of inertia is still the third of (\ref{27}):%
\begin{equation}
\dot{Y}=YB^{T}+BY\text{ \ \ , \ \ }B=\overset{\cdot }{G}G^{-1}.  \label{48}
\end{equation}

The evolution equations for $x$, $G$ and $H$ can be formally adapted from
the first, second and fourth of (\ref{27}):%
\begin{equation}
\begin{array}{c}
\rho \ddot{x}=\hat{f}, \\ 
\rho \left( \dot{K}-BK-H\right) =\hat{M}-\hat{A}\text{ \ \ , \ \ }K=YB^{T},
\\ 
\rho \left( \dot{H}+BH+HB^{T}\right) =\overset{\smallfrown }{S}-\hat{Z}.%
\end{array}
\label{49}
\end{equation}

The sources per unit volume of momentum, moment of momentum and stir need be
still specified. The choice is most difficult because those sources might
not be necessarily of purely local character; at least weakly non-local
effects may be expected. Thus a vast field of inquiry is opened, possibly,
even necessarily, involving concepts recalled in Section 5; below reference
is made only to very special choices advanced in important contexts. They
all presume that the gross local shape and orientation be invariant: $B=0$; $%
Y$, constant. This presumption implies the irrelevance of equation (\ref{48}%
) and the dropping of the second equation in system (\ref{41}), thus
implying that $A=\rho H$; besides, for $\hat{f}$, the standard Cauchy form
is postulated%
\begin{equation*}
\hat{f}=\rho b+div\Sigma ,
\end{equation*}%
requiring a volume action $b$ and a Cauchy stress $\Sigma $; finally, for $%
\Sigma $, the special constitutive relation $\Sigma =\rho H$ is accepted.
The choices of $\overset{\smallfrown }{S}$ and $\hat{Z}$ remain open and
again an extended Cauchy type Ansatz is called upon with a volume stir and a
shuffle flux expressed with the help of a hyperstress; for both stir and
hyperstress appropriate constitutive relations are finally chosen.

All these special choices seem to be satisfactory within the particular
contexts where they are advanced. But perhaps some general rules for $\hat{f}
$, $\hat{M}$, $\hat{A}$, $\overset{\smallfrown }{S}$ and $\hat{Z}$ should
rather be sought and corresponding fairly general subchapters of continuum
mechanics exhaustively written.

\ \ \ \ \ \ 

\textbf{Acknowledgement}. This research was supported by the Italian
M.I.U.R. through the project "Modelli Matematici per la Scienza dei
Materiali".

\section{References}

\begin{enumerate}
\item Capriz, G. (1999), Elementary preamble to a theory of kinetic
continua, in \emph{Atti del Convegno "La matematica nelle scienze della vita
e applicazioni"}, Bologna, 17-31.

\item P\"{o}schel, T. and Luding, S. (edts), \emph{Granular gases},
Springer-Verlag, Berlin, 2001.

\item Truesdell, C. (1955), Hypo-elasticity, \emph{J. Rational Mech. Anal.}, 
\textbf{4}, 83-133.

\item Grad, H. (1949), On the kinetic theory of rarefied gases, \emph{Arch.
Rational Mech. Anal.}, \textbf{2}, 331-407.

\item M\"{u}ller, I. and Ruggeri, T., \emph{Rational extended thermodynamics}%
, Springer-Verlag, New York, 1998.

\item Jenkins, J. T. and Richman, M. W. (1985), Grad's 13 moments system for
a dense gas of inelastic spheres, \emph{Arch. Rational Mech. Anal.}, \textbf{%
87}, 355-377.

\item Cohen, H. and Muncaster, R. G., \emph{The theory of pseudo-rigid bodies%
}, Springer-Verlag, Berlin, 1988.

\item Lewis, R. M. (1960), Measure-theoretic foundations of statistical
mechanics, \emph{Arch. Rational Mech. Anal.}, \textbf{5}, 355-381.

\item Goldreich, P. and Tremain, S. (1978), Velocity dispersion in Saturn's
rings, \emph{Icarus}, \textbf{34}, 227.
\end{enumerate}

\end{document}